\documentclass[openacc]{rstransa}
\usepackage{multicol}
\usepackage{booktabs}

\begin{document}

\title{Efficient supersonic flows through high-order guided equilibrium with lattice Boltzmann}

\author{
    Jonas Latt$^{1,2}$, Christophe Coreixas$^{1}$, Jo\"el Beny$^{1}$, Andrea Parmigiani$^{2}$}

\address{$^{1}$Department of Computer Science, University of Geneva, 1204 Geneva, Switzerland\\
$^{2}$FlowKit-Numeca Group Ltd, Route d’Oron 2, 1010 Lausanne, Switzerland}

\subject{computational physics, fluid mechanics}

\keywords{LBM, CFD, compressible, supersonic.}

\corres{Jonas Latt\\
\email{jonas.latt@unige.ch}}

\begin{abstract}
    A double-distribution-function based lattice Boltzmann method (DDF-LBM) is proposed for the simulation of polyatomic gases in the supersonic regime. The model relies on an extended equilibrium state that is constructed to reproduce 
    the first 13 moments of the Maxwell-Boltzmann distribution exactly. This extends the validity of the standard 5-constraint (mass, momentum and energy) approach and leads to the correct simulation of thermal, compressible flows with only 39 discrete velocities in 3D. The stability of this BGK-LBM is reinforced by relying on Knudsen-number-dependent relaxation times that are computed analytically.  Hence, high-Reynolds number, supersonic flows can be simulated in an efficient and elegant manner. While the 1D Riemann problem shows the ability of the proposed approach to handle discontinuities in the zero-viscosity limit, the simulation of the flow past a NACA0012 airfoil (Mach number $\mathrm{Ma}=1.5$, Reynolds number $\mathrm{Re=10^4}$) confirms the excellent behavior of this model in a low-viscosity and supersonic regime. The proposed model is substantially more efficient than the previous 5-moment D3Q343 DDF-LBM and opens up a whole new world of compressible flow applications that can be realistically tackled with a purely LB approach.
\end{abstract}
\maketitle

\section{Introduction}
The lattice Boltzmann method (LBM) is a popular numerical scheme capable of computing solutions of the Boltzmann equation (BE) in a regime of small deviations from the local equilibrium state~\cite{he1997priori,shan1998discretization}. It offers
a pathway to recover the solutions of the incompressible Navier-Stokes or compressible Navier-Stokes-Fourier equations~\cite{guo2013lattice,kruger_lattice_2017,succi_lattice_2018}. It can therefore be used as an alternative to classical solvers in the field of Computational Fluid Dynamics and has gained traction notably in areas with complex and coupled physics. Examples include multi-phase~\cite{leclaire_generalized_2017,parmigiani_lattice_2013} or particulate flows~\cite{thorimbert_lattice_2018}, or biomedical applications~\cite{li_application_2018}. A review of the method is for example provided in~\cite{chen_lattice_1998,kruger_lattice_2017,succi_lattice_2018}.
In the BE, the statistical behavior of gas molecules is described by the continuous velocity distribution function $f(\bm{x}, \bm{\xi}, t)$ that depends on the spatial position $\bm x$, the molecular velocity $\bm\xi$ and time $t$~\cite{huang_statistical_1987}. In the LB scheme, the space of molecular velocities is discretized: the velocity distribution function is replaced by a discrete set of $V$ populations $f_i(\bm{x}, t)$, $i=0\cdots V-1$ standing for the statistics of molecules at discrete velocities $\bm{\xi}_i$. Increasing the number of discrete velocities, and thus populations, typically allows to extend the physical range of validity of the scheme~\cite{shan_kinetic_2006}.

Although the LBM describes intrinsically compressible fluids, it has arguably achieved its most striking successes in the realm of incompressible fluid flow~\cite{kruger_lattice_2017,succi_lattice_2018} and has until recently struggled to establish itself as a realistic alternative for the simulation of compressible and/or high Mach number flows~\cite{guo_lattice_2013}. While simulating fluid flow may seem as easy as adding discrete velocities to the LBM scheme, these so-called multi-speed approaches rapidly lead to schemes with a large number of velocities, which translate in an impractically large number of degrees of freedom per mesh cell. In addition to that, these models are usually limited by numerical instabilities~\cite{siebert_lattice_2008}, except if the numerical scheme is modified~\cite{guo_lattice_2013}, or if a more sophisticated collision model is adopted~\cite{coreixas_recursive_2017,li_temperature-scaled_2019}.

As a popular workaround, the momentum and energy equations can be split and solved separately, as they individually require a substantially smaller set of velocities~\cite{li_coupled_2007,feng_compressible_2016}. Further computational efficiency is achieved by solving the energy equation with a traditional finite-difference, finite-volume, or finite-element scheme~\cite{nie_lattice-boltzmann_2009,fares_validation_2014}. These solutions introduce however new problems of their own making in the shape of coupling instabilities, that can be reduced through a careful choice of the coupling methodology~\cite{renard2019improved}.

Another interesting path has been taken by the community of Discrete Velocity Models (DVMs) to derive kinetic models devoted to the simulation of rarefied gas flows~\cite{gatignol1975theorie,charrier1998discrete,mieussens_discrete_2000,mieussens2001convergence,dubroca2001conservative,mieussens2004numerical,titarev2012efficient,baranger2014locally}, even though it is not restricted to this application field\cite{dubroca1999theoretical,charrier2004discrete}. Similarly to LBMs, the DVM also solves a discrete velocity BE. But contrary to LBMs, it is based on more complex numerical discretizations of this set of equations (finite-volume, finite-element, etc), and it relies on discrete equilibrium functions in the form of an exponential that are derived through the minimum-entropy principle~\cite{kogan_derivation_1965,levermore1996moment,presse_principles_2013}. 
In the context of rarefied gas flow, low-order DVMs compute this exponential equilibrium on each cell and at each time step by solving a set of non-linear equations, subject to only five constraints (mass, momentum, and energy conservation~\cite{mieussens_discrete_2000,dubroca2001conservative,mieussens2004numerical}), whereas high-order models can incorporate more constraints to increase their validity range~\cite{charrier1998discrete,titarev2012efficient}, as proposed by Levermore in his seminal work on the derivation of high-order closure of kinetic theories~\cite{levermore1996moment}. 
Existence, uniqueness and positiveness of solutions to this problem, by a root-finding algorithm like Newton-Raphson, has been extensively discussed by authors of the DVM. Best practices were also provided to accelerate the convergence and increase the robustness of the root-finding algorithm~\cite{mieussens_discrete_2000,titarev2012efficient}. Nevertheless, DVMs are computationally expensive due to the great number of discrete velocities --which usually exceeds a thousand in 3D-- to correctly recover the correct shape of populations in out-of-equilibrium regions of the simulation domain.

More recently, the 5-constraint (guided) equilibrium~\cite{mieussens_discrete_2000} has been (re)introduced in the context of compressible LBMs by Frapolli et al.~\cite{frapolli_entropic_2015,frapolli_entropic_2016}. Among DDF-LBMs based on the collide-and-stream algorithm, this is to our knowledge one of the most promising propositions in the current literature capable of achieving good numerical stability for non-trivial, transonic and supersonic flows. As for DVMs based on the 5-constraint equilibrium, the computational cost of this methodology is very high, as it typically requires two populations composed of 343 velocities each to recover the correct macroscopic behavior in the supersonic regime. As compared to the conventional 19 or 27 velocities needed to simulate weakly compressible and athermal flows, such an approach rapidly becomes prohibitive and faces severe difficulties to simulate realistic configurations in an acceptable time frame, or using an acceptable amount of hardware resources. Indeed, grid meshes composed of tens, or even hundreds of millions of points can be required to achieve the high accuracy required at an industrial level. Finally, the approach relies on the entropic collision model that requires minimizing the H-functional, on every single grid point and at every time iteration, in order to get more stable simulations. This eventually leads to a further non-negligible overhead.

In this publication, we extend the above compressible LBM by increasing the number of constraints used to compute the exponential equilibrium to 13, with the aim to increase its validity range and robustness. This follows an idea proposed in the context of DVMs by Charrier et al.~\cite{charrier1998discrete}, and also evoked in the PhD manuscript of Frapolli~\cite{frapolli_entropic_2017}. With 13 constraints, the equilibrium distribution matches the moments of the Maxwell-Boltzmann distribution not only for the conserved moments, as does the 5-moment approach, but also for higher-order, non-conserved moments. In this manner, the behavior of the macroscopic flow variables, which are related to the moments of the particle populations, can be adequately represented using as few as 39 discrete velocities in 3D, by putting more effort on the equilibrium instead of the lattice (Section~\ref{sec:model}). This is explained by the fact that the desired moments of the distribution are strictly enforced with the help of a numerical solver. Hence, they no longer depend on the choice of a sophisticated discrete velocity stencil, or at least, the dependency is highly reduced. 

Compared to a 5-constraint approach, the memory requirements are thus reduced by an order of magnitude, and the computational cost is diminished correspondingly. In contrast to the entropic collision model, we prefer to rely on BGK operators~\cite{BHATNAGAR_PR_94_1954} that are stabilized by identifying areas of the simulation domain where the departure from equilibrium is high. In doing so, relaxation times can be adjusted to damp high-order modes, which efficiently leads to stable and accurate simulation of supersonic flows with discontinuities. Consequently, the proposed model combines the elegance of a fully LB model for compressible flows with a level of robustness and efficiency.

The rest of the paper reads as follows. First, the derivation of discrete equilibrium based on the maximum entropy principle is recalled in the context of fluid mechanics, with a particular emphasis on the model used in this first study (Section~\ref{sec:model}). The numerical discretization and the kinetic sensor are presented in Section~\ref{sec:numerical-model} 
Numerical tests shown in this article confirm that the model exhibits both a good stability and accuracy with this reduced velocity set (Section~\ref{sec:numerical-tests}). The computational expense of the method is measured on both CPU and GPU hardware in Section~\ref{sec:efficiency}. Conclusions and perspectives are eventually provided in Section~\ref{sec:conclusion}.

\section{The model \label{sec:model}}
The Maxwell-Boltzmann distribution
\begin{equation}
f^{eq}=\frac{\rho}{(2\pi T)^{D/2}}\exp\left(\frac{(\bm{\xi}-\bm{u})^2}{2T}\right),
\end{equation}
describes an equilibrium state and yields an exact solution of the Boltzmann equation. The macroscopic density $\rho$, velocity $\bm{u}$ and temperature $T$ are equal to or directly linked with velocity moments of the probability distribution function. A macroscopic variable is a collision invariant, and thus a conserved quantity, if the corresponding moment of the distribution function $f$ and of the equilibrium state $f^{eq}$ are equal. It is easily verified that for the continuum Maxwell-Boltzmann distribution, density, velocity, and total energy are conserved quantities.

In the discrete LB method, the equilibrium populations $f_i^{eq}$ are a mere approximation of the Maxwell-Boltzmann distribution, but most LB models guarantee the correct definition of conserved macroscopic quantities after the velocity space discretization. To achieve this goal, the most common methodology is to rely on Gauss-Hermite quadrature~\cite{shan_kinetic_2006}. In this case, the match between moments and macroscopic variable is nothing else than a statement of orthogonality between Hermite polynomials. However, the number of discrete velocities required to express appropriate orthogonality relations and include effects of energy conservation can be prohibitively large (at least 100 velocities in 3D). As an alternative, other authors proposed to manually enforce conservation laws by constantly recomputing the equilibrium by a root-finding procedure subject to appropriate constraints~\cite{frapolli_entropic_2015}.

Beyond conserved quantities, it is however necessary for the moments of the discrete equilibrium to yield the same value as the corresponding moments of the Maxwell-Boltzmann distribution in order to recover the desired physical behavior. We now provide a short reasoning for this argument, and compute the order of the moments that need to be recovered exactly to produce Navier-Stokes-Fourier (NSF) level physics. The compressible NSF equations read 
\begin{equation}
\begin{array}{c}
\partial_t \rho + \partial_{\chi}(\rho u_{\chi})=0,\\[0.1cm]
\partial_t (\rho u_{\alpha}) + \partial_{\beta}(\rho u_{\alpha}u_{\beta} + p \delta_{\alpha\beta})=\partial_{\beta}(\Pi_{\alpha\beta}),\\[0.1cm]
\partial_t (\rho E) + \partial_{\alpha}((\rho E +p)u_{\alpha})=\partial_{\alpha}(\Phi_{\alpha}),
\end{array}
\label{eq:NSF}
\end{equation}
where the index repetition implies the Einstein summation rule. The viscous stress tensor $\Pi_{\alpha\beta}$ is defined as 
$$\Pi_{\alpha\beta} = \mu (S_{\alpha\beta}-\tfrac{2}{D}\partial_{\chi}u_{\chi}\delta_{\alpha\beta}) + \mu_b\partial_{\chi}u_{\chi}\delta_{\alpha\beta},$$
with $S_{\alpha\beta}=\partial_{\alpha}u_{\beta}+\partial_{\beta}u_{\alpha}$, and $\mu$ and $\mu_b=(2/D-1/C_v)\mu$ being the dynamic and bulk viscosity respectively. $D$ is the number of physical dimensions, $C_v=1/(\gamma-1)$ is the heat capacity at constant volume,$\gamma$ is the heat capacity ratio, and $E$ is the total energy. Dissipative effects in the total energy equation are gathered in the term
$$\Phi_{\alpha} = -\lambda\partial_{\alpha}T + u_{\beta}\Pi_{\alpha\beta},$$
which accounts for both the Fourier heat flux ($\lambda$ is the heat conductivity), and the viscous heat dissipation. In case of a calorically perfect gas, this system is closed by the following equation of state: $E=\tfrac{1}{2}u^2+C_v T$.

The moments of the Maxwell-Boltzmann distribution can be written as
\begin{equation}
M^{\mathrm{MB}}_{pqr}=\displaystyle{\int \xi_x^p\xi_y^q\xi_z^r f^{eq}\,d\bm{\xi}}.
\label{eq:ContinuousMoments}
\end{equation}
For comparison with macroscopic equations, it is more convenient to write them in the following compact form:
\begin{align}
M^{\mathrm{MB}}_{0}&=\rho,\nonumber\\
M^{\mathrm{MB}}_{1,{\alpha}}&=\rho u_{\alpha},\nonumber\\
M^{\mathrm{MB}}_{2,{\alpha\beta}}&=\rho u_{\alpha}u_{\beta}+ \rho T\delta_{\alpha\beta},\label{eq:EqMomentsTensor}
\\
M^{\mathrm{MB}}_{3,{\alpha\beta\gamma}}&=\rho u_{\alpha}u_{\beta}u_{\gamma}+\rho T[u_{\alpha}\delta_{\beta\gamma}]_{{cyc}},\nonumber\\
M^{\mathrm{MB}}_{4,{\alpha\beta\gamma\chi}}&=\rho u_{\alpha}u_{\beta}u_{\gamma}u_{\chi} + \rho T[u_{\alpha}u_{\beta}\delta_{\gamma\chi}]_{{cyc}} + \rho T^2[\delta_{\alpha\beta}\delta_{\gamma\chi}]_{{cyc}},\nonumber
\end{align}
where $\alpha$, $\beta$, $\gamma$, $\chi$ represent space coordinates ($x$, $y$ or $z$). The subscript ${{cyc}}$ labels a cyclic permutation without repetition. As an example,
$$T^2[\delta_{\alpha\beta}\delta_{\gamma\chi}]_{\mathrm{cyc}}=T^2(\delta_{\alpha\beta}\delta_{\gamma\chi} + \delta_{\alpha\gamma}\delta_{\beta\chi} + \delta_{\alpha\chi}\delta_{\beta\gamma}).$$
Now, the NSF equations are easily rewritten in terms of equilibrium moments as follows:
\begin{align}
\label{eq:EulerMomentConstraints}
    \partial_t (M^{\mathrm{MB}}_{0}) + \partial_{\beta}(M^{\mathrm{MB}}_{1,\beta})&=0,\\
    \partial_t (M^{\mathrm{MB}}_{1,\alpha}) + \partial_{\beta}(M^{\mathrm{MB}}_{2\alpha\beta}) &\propto \partial_t (M^{\mathrm{MB}}_{2\alpha\beta}) + \partial_{\gamma} (M^{\mathrm{MB}}_{3,\alpha\beta\gamma}),\\
    \partial_t (M^{\mathrm{MB}}_{2\alpha\alpha}) + \partial_{\beta}(M^{\mathrm{MB}}_{3,\alpha\alpha\beta}) &\propto \partial_t (M^{\mathrm{MB}}_{3,\alpha\chi\chi}) + \partial_{\beta} (M^{\mathrm{MB}}_{4,\alpha\beta\chi\chi}).
\end{align}
Here, the diffusive RHS terms have been related to equilibrium moments through the Chapman-Enskog expansion, where the time derivatives are usually replaced using Euler-level equations for $M^{\mathrm{MB}}_{2,\alpha\beta}$ and $M^{\mathrm{MB}}_{3,\alpha\gamma\gamma}$. From this, it is clear that moments of $f^{eq}$ up to $n=4$ are necessary to recover NSF equations~(\ref{eq:NSF}).\\
Consequently, the matching conditions between discrete and continuous moments read:
\begin{equation}
\sum_i f^{eq}_i \xi_x^p\xi_y^q\xi_z^r = M_{pqr}^{\mathrm{MB}}.
\end{equation}
with $p+q+r=n$. These conditions are enforced using the following (exponential) discrete equilibrium:
\begin{equation}\label{eq:exponential}
 f_i^{eq} = \rho \exp[-(1+\textstyle{\sum_{p,q,r}}\lambda_{M_{pqr}^{\mathrm{MB}}}\xi_x^p\xi_y^q\xi_z^r)],
\end{equation}
where $\lambda_{M_{pqr}^{\mathrm{MB}}}$ are Lagrange multipliers whose purpose is to match the following constraints:
\begin{equation}
G_{pqr}=\sum_i f_i^{eq} \xi_x^p\xi_y^q\xi_z^r - M_{pqr}^{\mathrm{MB}}=0\label{eq:constraints},
\end{equation}
This particular form of the equilibrium~(\ref{eq:exponential}) is obtained via the principle of maximum entropy, considering the following $H$-function 
$H=\sum_i f_i \ln{f_i}$~\cite{kogan_derivation_1965,levermore1996moment,presse_principles_2013}, and under the constraints~(\ref{eq:constraints}).\\
It is useful to note that contrarily to standard LB models, the above equilibrium does not explicitly introduce lattice weights. The latter originates from the quadrature used for the standard construction of the lattice and its corresponding polynomial equilibrium. Such a methodology is not carried anymore in the present context, hence the solution procedure is quadrature-free.

In case of the NSF equations, the moments should ideally be matched up to fourth order. In this first study, we restrict ourselves however to the first 13 moments, while forthcoming studies will consider higher-order moments (R26, R45, etc)~\cite{struchtrup_regularization_2003}. Here, the discrete equilibrium reads
\begin{equation}\label{eq:discrete-equilibrium}
    f_i^{eq} = \rho \exp{\left(\lambda_0 + {\lambda}_{1,\alpha}{\bf\xi}_{i,\alpha} + {\lambda}_{2,\alpha\beta}{\bf\xi}_{i,\alpha}{\bf\xi}_{i,\beta} + {\lambda}_{3,\alpha}{\bf\xi}_{i,\alpha}\left|{\bf\xi}_{i}\right|^2\right)}.
\end{equation}	
Here, ${\lambda}_{2}$ is a symmetric tensor with 6 independent variables. Thus, the equilibrium depends only on 13 unknown variables, which are matched with the following 13 constraints:
\begin{align}\label{eq:13constraints}
M_{0}^{eq} &= \sum_i f_i^{eq} = \rho\\
M_{1,\alpha}^{eq} &= \sum_i f_i^{eq} \xi_{i\alpha} = \rho u_\alpha\\
M_{2,\alpha\beta}^{eq} &= \sum_i f_i^{eq} \xi_{i\alpha}\xi_{i\beta} = \rho u_\alpha u_\beta + \rho T\delta_{\alpha\beta}\\
M_{3,\alpha\beta\beta}^{eq} &= \sum_i f_i^{eq} \xi_{i\alpha} \left|\xi_{i}\right|^2 = \left(2\rho E^{tr} + 2 \rho T\right)u_\alpha
\end{align}
where the total translational energy $E^{tr}$ reads $E^{tr} = D T + \rho u^2$. These equations express the conservation of mass and momentum, and enforce a desired form for the pressure tensor (which includes the conservation law for the translational energy) and the contracted heat flux tensor. 
To further extend the above approach to polyatomic gases, one can rely on a second set of populations $g_i$ described by the following equilibrium~\cite{rykov_model_1975,nie_thermal_2008}
\begin{equation}
g_i^{eq} =(2C_v-D)T f_i^{eq},
\label{eq:discrete-equilibrium-poly}
\end{equation}
that is built on top of $f_i^{eq}$.

\section{Numerical method \label{sec:numerical-model}}
The above model is based on the collide-and-stream algorithm in a space of discrete velocities extending to up to three neighbors. We selected the D3Q39 lattice, which is proposed in~\cite{shan_kinetic_2006} under the name $E_{3,7}^{39}$, and contains the following velocities (FS denotes a fully symmetric set of points): $(0,0,0)$, $(1,0,0)_{FS}$, $(\pm 1,\pm 1,\pm 1)$, $(2,0,0)_{FS}$, $(2,2,0)_{FS}$, $(3,0,0)_{FS}$. For the collision step, each cell first computes the macroscopic variables ($\rho$, $\bm u$, $T$ and $E^{tr}$) and evaluates the moments, Eq.~(\ref{eq:13constraints}). Then, the 13 unknown variables $\lambda_i$ are computed by plugging Eq.~(\ref{eq:discrete-equilibrium}) into equations~(\ref{eq:constraints}) and solving the system (we used a standard multivariate Newton-Raphson root solver~\cite{press_numerical_2007}). Once the $f_i^{eq}$ are known, $g_i^{eq}$ is computed using Eq.~(\ref{eq:discrete-equilibrium-poly}). Adopting the BGK collision model, post-streaming populations are finally computed using the standard collide-and-stream algorithm, formulated in lattice units:
\begin{equation}
    h_{i}\left(\boldsymbol{x}+\boldsymbol\xi_{i}, t + 1\right) = h_{i}\left(\boldsymbol{x}, t\right) \,-\, \dfrac{1}{\tau_h} \left(h_{i}-h_{i}^{(eq)}\right)\left(\boldsymbol{x}, t\right),
\end{equation}
where $h$ is the population, representing $h=f$ or $h=g$.
In addition, the Prandtl number could be adjusted using either the quasi-equilibrium~\cite{frapolli_entropic_2016}, or a more standard approach~\cite{li_temperature-scaled_2019}. This was not implemented for this first study, which limits the simulations to $\mathrm{Pr}=1$. Nevertheless, in the low-viscosity regime considered hereafter, it is expected that the error induced by such an approximation is negligible.

In case better stability is required, the relaxation time is made Knudsen-number dependent. This is done by locally computing
\begin{equation}
\epsilon = \dfrac{1}{V}\sum_{i=0}^{V-1} \dfrac{\vert f_i-f_i^{eq}\vert}{f_i^{eq}},
\label{eq:DeltaAvg}
\end{equation}
where $\epsilon \approx \mathrm{Kn}$ according to the Chapman-Enskog expansion at the Navier-Stokes Fourier level~\cite{chapman_mathematical_1990}. For large departures from the equilibrium state, such that $\epsilon > 10^{-2}$, the continuum limit assumption does not hold anymore, and the recovery of the macroscopic behavior of interest is then at risk. 
To prevent such an issue, one could refine the grid mesh depending on the value of $\epsilon$, as proposed by Thorimbert et al.~\cite{thorimbert_automatic_2019}. Here, we prefer to damp high-order contributions, as soon as they appear in under resolved areas of the simulation domain, by adjusting the relaxation time accordingly to the local departure from the equilibrium state:
$\tau(\epsilon)= \tau \alpha(\epsilon)$. 
As a simple proof of concept, $\alpha$ is chosen as a slowly increasing piecewise constant function of the local Knudsen number $\epsilon$ which, in the worst case, replaces the populations by equilibrium ($\epsilon \geq 1$):
\begin{equation}
\alpha = \left\{
\begin{array}{lll}
1, & &\epsilon < 10^{-2} \\
1.05, & \quad 10^{-2} \leq &\epsilon < 10^{-1}\\
1.35, & \quad 10^{-1} \leq &\epsilon < 1\\
1/\tau, & \quad  & \epsilon \geq 1
\end{array} 
\right.
\label{eq:alpha}
\end{equation}
where $1/\tau \approx 2$ in the context of high-Reynolds number flows, which ensures the well-posedness of $\alpha$. This methodology amounts to artificially increasing the viscosity in areas of strong departure from the equilibrium state, and as such, is similar to both a subgrid scale model and a shock sensor. This information could also be used to better control the relaxation of high-order moments, similarly to the KBC collision model~\cite{karlin_gibbs_2014}. Such an investigation will be presented in a future work. 

It is finally worth noting that our stabilization technique share the same philosophy as the so-called entropic filtering approach~\cite{BROWNLEE_PRE_74_2006,BROWNLEE_PRE_75_2007,BROWNLEE_PA_387_2008,gorban_enhancement_2014}. For the latter, the LB scheme is also stabilized by introducing an artificial dissipation term that depends on the local departure from equilibrium, which can be estimated through several entropic metrics. As an example, the following quadratic entropy estimate
$$
\Delta S = \sum_{i=0}^{V-1} \dfrac{( f_i-f_i^{eq})^2}{f_i^{eq}},
$$
was used in the comparative study by Gorban and Packwood~\cite{gorban_enhancement_2014} to dynamically compute either second- or higher-order relaxation times by making them proportional to $\Delta S$. The use of the $l_1$-norm 
\begin{equation}
\vert\vert \bm{f}-\bm{f^{eq}}\vert\vert_1 = \sum_{i=0}^{V-1} \vert f_i-f_i^{eq}\vert,
\label{eq:norm1}
\end{equation}
was also evoked by the latter authors, without being investigated in their work. Yet, the normalisation by $f_i^{eq}$ is missing from the $l_1$-norm~(\ref{eq:norm1}), and this prevent the proper evaluation of the departure from equilibrium in terms of Knudsen number. All of this confirms that our stabilization technique cannot be recast in terms of entropic filtering, but instead, it is an innovative way to improve the numerical stability of LBMs through kinetic considerations.

\section{Numerical tests}\label{sec:numerical-tests}
\begin{figure}[bp]
\begin{minipage}{.48\textwidth}
\includegraphics[width=\textwidth]{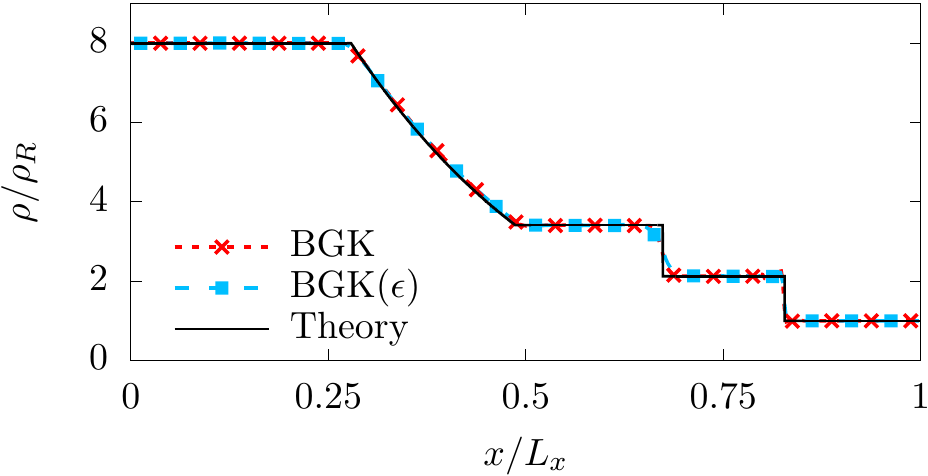}
\end{minipage}
\hfill
\begin{minipage}{.48\textwidth}
\hfill
\includegraphics[width=\textwidth]{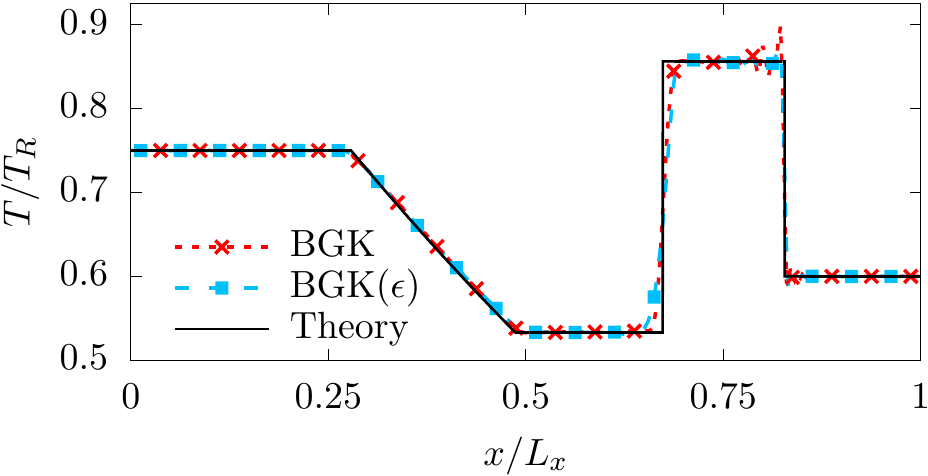}
\end{minipage}
    \caption{Sod shock tube for a diatomic gas (specific heat ratio $\gamma=7/5$). Results obtained with $L_x=400$ points using the standard BGK, and our version [$\mathrm{BGK }(\epsilon)$] are compared against theoretical curves, for $\tau_f=0.57$ and $0.5$
respectively. 
    }\label{fig:sod-shock}
\end{figure}

As a first validation, the generation and propagation of both rarefaction and shock waves is investigated through the simulation of a 1D Riemann problem, also known as Sod shock tube~\cite{sod_survey_1978}. The following initial setup is used:
$(\rho_L, T_L)=(8,10)$ and $(\rho_R, T_R)=(1,1)$.
Subscripts $L$ and $R$ stand for the left and the right states, respectively. The gas is considered at rest for both states, and the simulation domain is discretized along the $x$-direction using only $L_x=400$ grid points. 
Corresponding results are compiled in Fig.~\ref{fig:sod-shock} for both the standard BGK operator and our dynamic version based on Eq.~(\ref{eq:alpha}). They prove the ability of the proposed model to simulate the generation and propagation of rarefaction and shock waves in a diatomic gas. Simulations are stable for relatively low values of the relaxation time ($\tau_f=0.57$, $\mathrm{Pr}=1$), even with the standard BGK operator, while LBMs based on a polynomial discrete equilibrium usually require a large number of discrete velocities (37 in 2D) and more sophisticated collision models in order to achieve similar results~\cite{coreixas_recursive_2017}. Interestingly, the Knudsen number based relaxation times drastically improve the stability of the present model, allowing the simulation of 1D Riemann problem in the zero-viscosity limit ($\tau_f=0.5$, $\mathrm{Pr}=1$), with an accuracy that competes with LBMs coupled with shock capturing techniques~\cite{coreixas_high-order_2018}. 

The next validation test consists of the simulation of the flow past a NACA0012 airfoil in the supersonic regime, and for a relatively high Reynolds number ($\mathrm{Re}=10^4$). Before moving to the results, it is worth noting that the standard BGK-LBM based on the equilibrium~(\ref{eq:discrete-equilibrium}) led to stable simulations for Mach and Reynolds numbers up to $1.2$ and $7500$ respectively. These parameters are however insufficient for a comparison against data available in the literature~\cite{hafez_simulations_2007,frapolli_entropic_2016}, which are provided for $(\mathrm{Ma},\mathrm{Re})=(1.5,10^4)$. Hence, the dynamic relaxation time based on the $\alpha$ function~(\ref{eq:alpha}) was used to further extend the stability range of the present approach. The simulation domain is defined as $[n_x,n_y,n_z]=[8C,8C,1]$ with $C=350$ points, $n_j$ being the number of points in the $j$ direction ($j=x,y$ or $z$), and centered around the leading edge of the airfoil. While the freestream conditions are imposed on the left, top and bottom boundary conditions of the domain (by imposing $f_i=f_i^{eq}$ and $g_i=g_i^{eq}$), a (first-order) Neumann boundary condition is imposed at the (right) outlet. The no-slip boundary condition is imposed on the airfoil using the half-way bounce-back methodology~\cite{kruger_lattice_2017}.

\begin{figure}[tbp]
        \centering
\begin{minipage}{.48\textwidth}
\includegraphics[width=\textwidth]{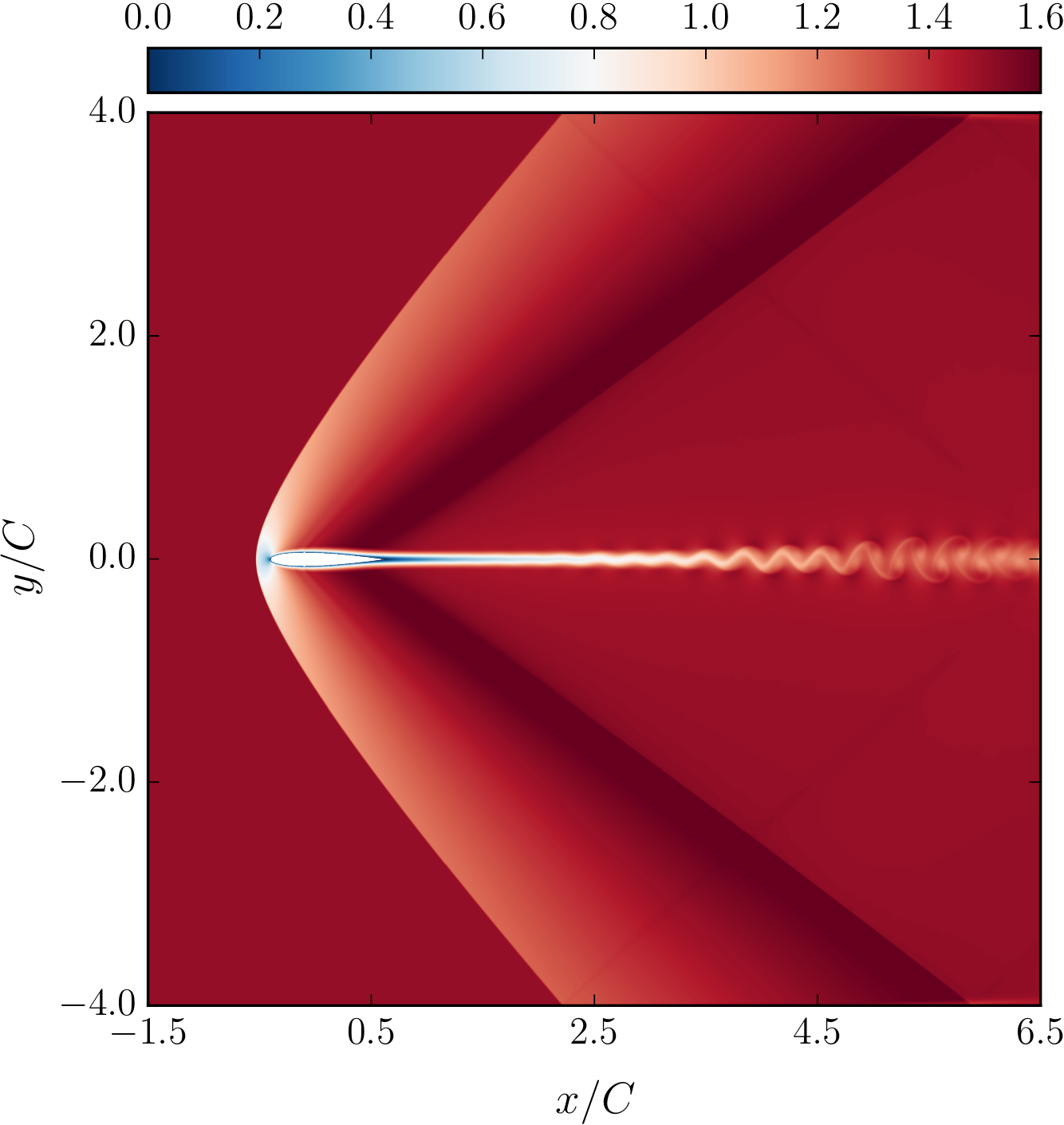}
\end{minipage}
\hfill
\begin{minipage}{.48\textwidth}
\includegraphics[width=\textwidth]{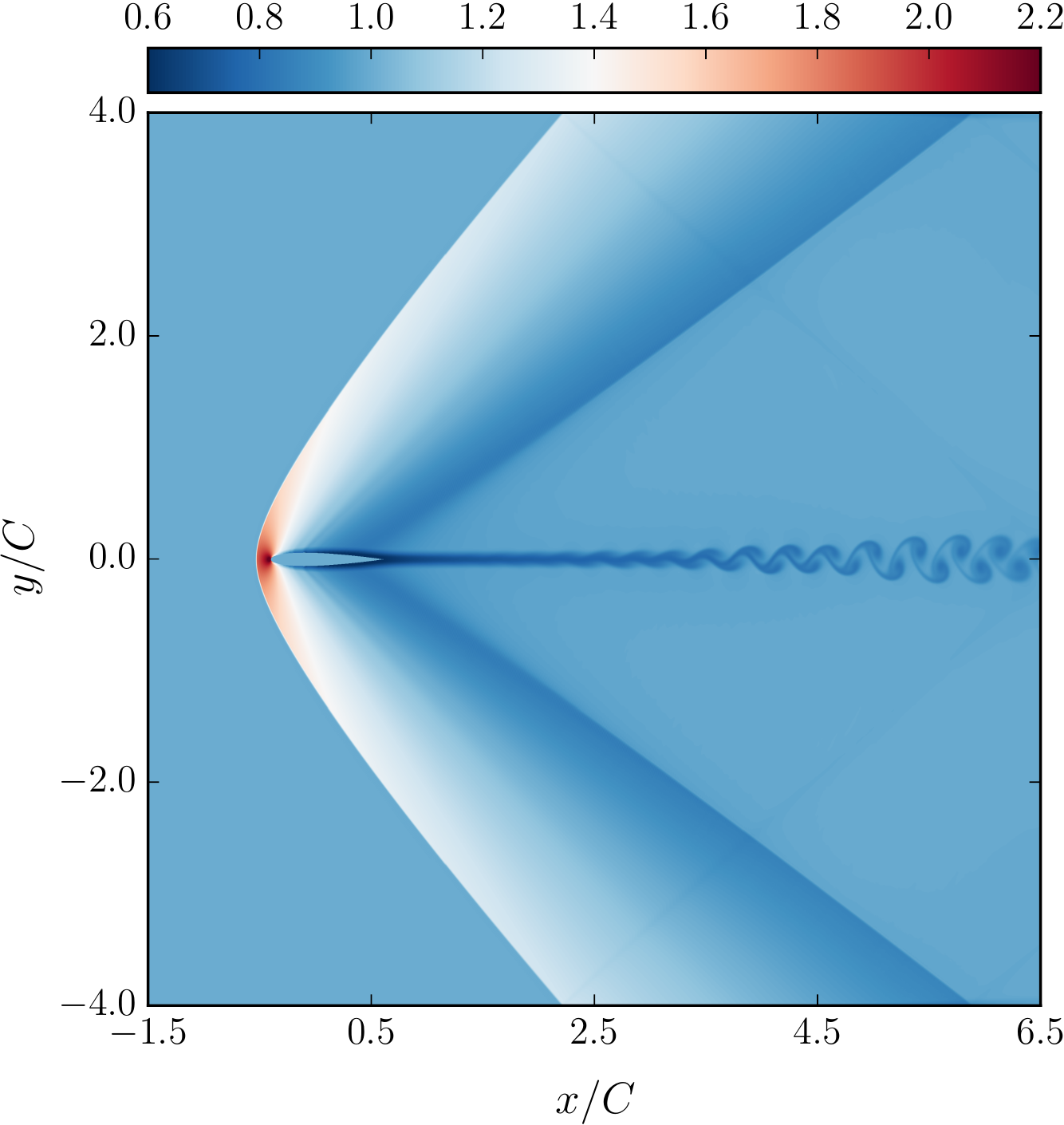}
\end{minipage}
\caption{NACA0012 airfoil at $\mathrm{Ma}=1.5$ and $\mathrm{Re}=10^4$ using $C=350$ points: local Mach number (left) and density (right) fields.}
    \label{fig:naca-mach-15}
\end{figure}

\begin{figure}[tbp]
        \centering
    \begin{minipage}{.48\textwidth}
    \includegraphics[width=\textwidth]{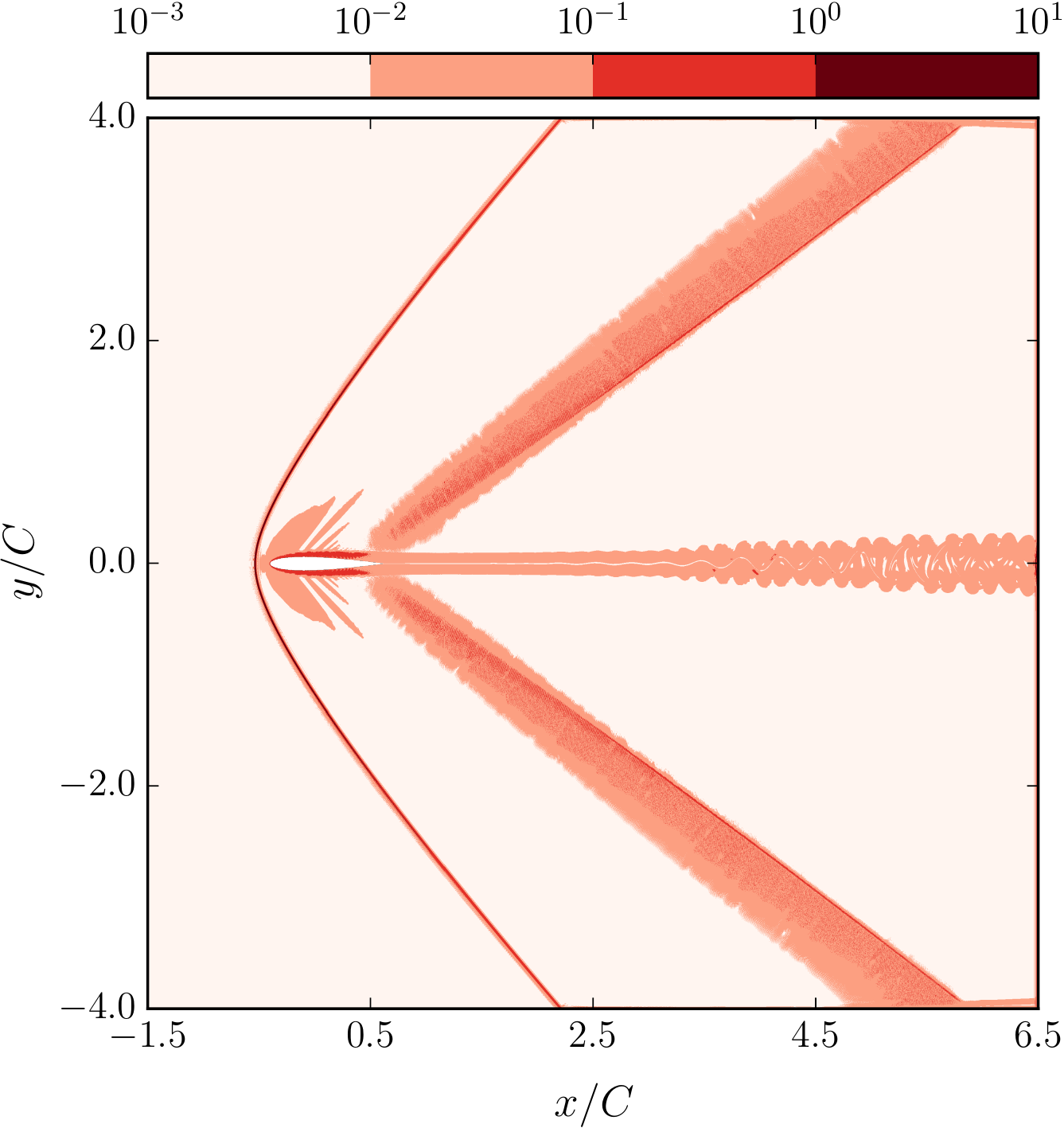}
    \end{minipage}
    \hfill
    \begin{minipage}{.48\textwidth}
    \includegraphics[width=1.05\textwidth]{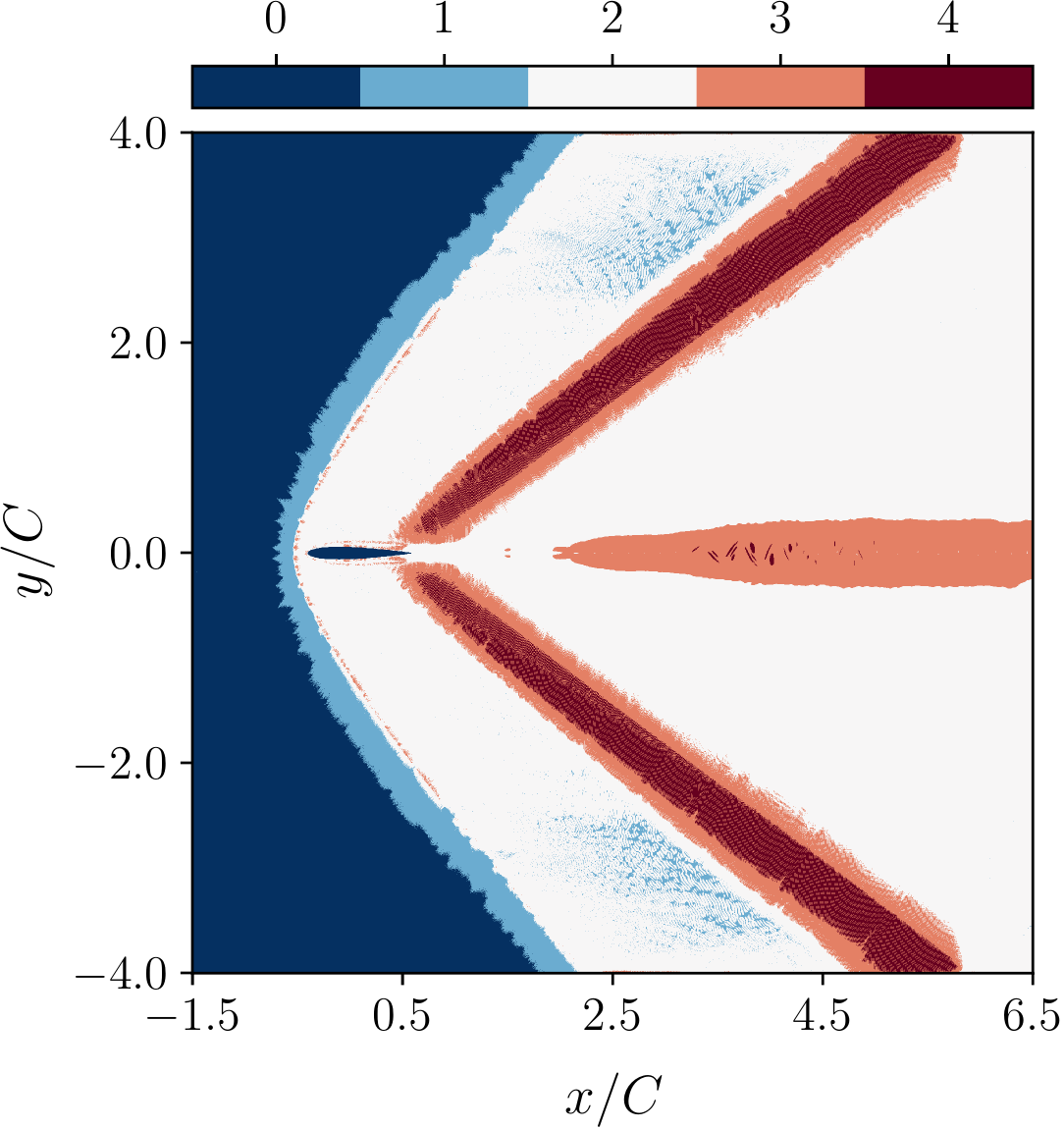}
    \end{minipage}
	\vspace*{-0.25cm}
    \caption{NACA0012 airfoil at $\mathrm{Ma}=1.5$ and $\mathrm{Re}=10^4$ using $C=350$ points: local Knudsen number $\epsilon$ (left), and number of iterations of the Newton-Raphson solver (right).}
    \label{fig:naca-knudsen}
\end{figure}

As a first insight on the physical phenomena related to this numerical validation, the local Mach number and density fields are plotted in Fig.~\ref{fig:naca-mach-15}. They highlights the main features of the flow: (1) primary strong bow shock upward the leading edge, (2) secondary weak shock close to the trailing edge, and (3) vortex shedding downward the airfoil. This qualitatively confirms the good behavior recovered thanks to the new equilibrium~(\ref{eq:discrete-equilibrium}), as compared to results published in~\cite{hafez_simulations_2007,frapolli_entropic_2016}.
To investigate the numerical properties of the present approach, the local Knudsen number $\epsilon$, as well as the number of Newton-Raphson iterations are plotted in Fig.~\ref{fig:naca-knudsen}. 
Interestingly, the kinetic sensor $\epsilon$ is only active close to the aforementioned features, which allows to properly classify them in terms of departure from the equilibrium. 
This observation serves as a justification for the fact that the adaptive BGK operator leads to stable simulations through a local increase of the kinematic viscosity adapted to the flow features: high increase for both strong and weak shocks, moderate close to the airfoil, low in its wake, none in the rest of the domain. 
This indicates that one of the possible outcomes of such a sensor would be a better control over areas of the simulation domain where shock-capturing techniques, and/or subgrid scale models should be activated. In addition, it is confirmed that the number of Newton-Raphson solver iterations required to impose the correct 13 moments is relatively low. In this example, it is limited to four at most and less than two on average, and is actually zero over a wide area, as the Lagrange multipliers inherited from the previous time step remain satisfying. It is also interesting to point out that the number of iterations follows the main flow features. Surprisingly, the highest number of iterations is not observed in the region where the highest departure from equilibrium is reported (bow shock), but instead, it is needed close to the secondary shock where the highest Mach number is encountered. This information indicates that a potential stability issue could arise from this part of the simulation domain. Hence, it is hypothesized that the number of iterations could be used to further activate last-resort stabilization techniques if needed be.


\begin{figure}
    \centering
    \includegraphics[width=.6\textwidth]{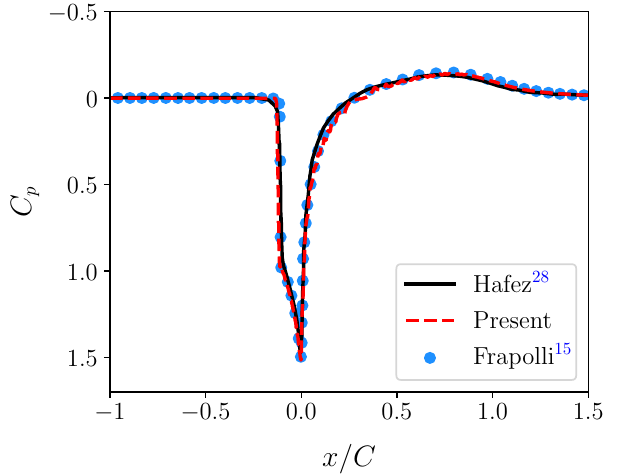}
    \caption{Comparison of $C_p$ distribution around the NACA0012 airfoil at $\mathrm{Ma}=1.5$ and $\mathrm{Re}=10^4$ using $C=350$ points.}
    \label{fig:cp}
\end{figure}

Quantitatively speaking, the comparison of pressure coefficient profiles ($C_p=(p-p_{\infty})/{0.5 \rho_{\infty} u^2_{\infty}}$) on figure~\ref{fig:cp} clearly shows that such a stabilization technique does not impact the level of accuracy reached by the present approach. More precisely, the width of the bow shock is fairly similar to reference solutions despite the equilibration of populations that is imposed in this particular region of the simulation domain. In addition, even if the bounce-back methodology leads to small oscillations close to the walls due to its inherent stair-case approximation, the parietal $C_p$ profile ($0\leq x/C \leq 1$) is also in agreement with reference studies. All in all, considering the relatively coarse grid mesh, as compared to~\cite{frapolli_entropic_2017} where $C=800$ grid points were used for the $343$-shifted-velocity $5$-constraints method, and the very simple no-slip boundary condition used in this study, these results are very encouraging.
It is eventually worth noting that stable simulations were obtained for $\mathrm{Re}=10^9$ in under resolved conditions ($C=200$ points). This confirms the excellent properties of our Knudsen number based stabilization technique.  

\section{Implementation and efficiency}\label{sec:efficiency}
Common LB schemes are characterized by a separation between a local collision step, which characterizes the physical properties of the model, and a generic, non-local streaming step. The collision step is usually carried out by manipulating the component of the populations that deviates from equilibrium, through a linear term, as in the straightforward BGK model~\cite{BHATNAGAR_PR_94_1954}, or a more general expression, as in multiple-relaxation-time approaches~\cite{KRUGER_Book_2017}. 
A crucial importance is therefore attributed to the explicit calculation of the equilibrium term, as a function of the macroscopic variables density, velocity, and temperature. 
As a matter of fact, the expression of the equilibrium distribution can be considered to entirely determine the physics expressed by the model~\cite{coreixas2019comprehensive}, if and only if the different non-equilibrium contributions are relaxed to recover the correct transport coefficients of the macroscopic equations of interest (see Ref.~\cite{SHAN_PRE_100_2019} for a systematic approach in the context of fluid mechanics). 

While LB models frequently approximate the equilibrium through an explicitly constructed polynomial expression~\cite{kruger_lattice_2017,succi_lattice_2018}, the present paper follows the path of DVMs and adopts a computationally more expensive approach of exponential-of-polynomial expression with iteratively computed coefficients~\cite{mieussens_discrete_2000,titarev2012efficient}. In details, the equilibrium is computed through the following process:

\begin{tabular}{@{}p{0.03\textwidth}@{}p{0.03\textwidth}@{}p{.37\textwidth}p{.3\textwidth}p{.1\textwidth}}
&\multicolumn{2}{@{}p{0.4\textwidth}@{}}{\textbf{Step}} & \textbf{Computation} & \textbf{Number of scalars}\\
--&\multicolumn{2}{@{}p{0.4\textwidth}@{}}{Compute macroscopic variables density, velocity, and temperature} & $f\quad \mapsto\quad (\rho, {\bm u}, T)$ & 5\\
--&\multicolumn{2}{@{}p{0.4\textwidth}@{}}{Compute moments of exact Maxwell-Boltzmann distribution} & $(\rho, {\bm u}, T)\quad \mapsto\quad M_{pqr}^{\mathrm{MB}}$ & 13\\
--&\multicolumn{2}{@{}p{0.4\textwidth}@{}}{Initialize Lagrange multipliers} & $\lambda_{M_{pqr}^{\mathrm{MB}}}(t-1)\quad \mapsto\quad \lambda^{(0)}(t)$ & 13\\
--&\multicolumn{4}{@{}l@{}}{Repeat until norm of error, Eq.~\ref{eq:constraints} is less than $10^{-12}$}\\
&--&Compute equilibrium distribution for current estimate of $\lambda$ (Eq.~\ref{eq:exponential}) & $\lambda^{(k)}\quad \mapsto\quad f_i^{\mathrm{eq},(k)}$ & 39\\
&--&Compute moments for current estimate of $\lambda$  and evaluate the constraint (Eq.~\ref{eq:constraints}) & ($f_i^{\mathrm{eq},(k)}, M_{pqr}^{\mathrm{MB}})\quad \mapsto\quad G_{pqr}$ & 13\\
&--& Compute Jacobian of $G_{pqr}$ with respect to $\lambda$ & $f_i^{\mathrm{eq},(k)}\quad \mapsto\quad J$ & $13 \times 13$\\
&--& Compute inverse of $J$ & $J\quad \mapsto\quad J^{-1}$ & $13 \times 13$\\
&--& Take $\lambda$ to the next iteration & $(\lambda^{(k)}, J^{-1})\quad \mapsto\quad \lambda^{(k+1)}$ & 13
\end{tabular}

The computational expense of this algorithm is dominated by the inversion of the Jacobian $J$. It can be mentioned that while the construction of $J$ itself is costly as well, the number of terms to be computed can be reduced from a total of $13 \cdot 13 = 169$ terms to only $48$, due to sparsity and redundancy of $J$. It is interesting to point out that in spite of the added complexity, the proposed model remains entirely local. It therefore increases the number of arithmetic operations to be executed per cycle, but adds little to the total number of memory accesses, which favorably matches the expectations of modern computing expectations. In particular, such properties can be considered in favor of the use of hardware accelerators.

For the purposes of this article, the model was implemented as a self-standing CPU code, written in the C++ language, and a self-standing GPU code written in CUDA. 
These codes cannot be compared in a straightforward manner to 5-moment entropic models, as these require the use of a D3Q343 lattice to achieve supersonic results as those presented in Section~\ref{sec:numerical-tests}. 
Comparisons with models based on a polynomial equilibrium are just as difficult, as the current literature provides no evidence that such an approach could allow to simulate non-trivial supersonic flows, except by introducing a substantially more complex collision model~\cite{coreixas_recursive_2017,coreixas_high-order_2018}. 
A comparison is nevertheless presented with a BGK model with a fourth-order polynomial equilibrium, extended to fourth order with respect to the Mach number. While the numbers cannot be taken as a competitive comparison (the polynomial model cannot solve the tests of Section~\ref{sec:numerical-tests} with the D3Q39 lattice~\cite{shan_kinetic_2006,SHAN_JCS_17_2016}), they provide a general understanding of overall performance of our model. 

The polynomial model is executed on the D3Q39 lattice and uses a double set of population to mimic the double-population approach of our model to vary the adiabatic expansion coefficient~\cite{rykov_model_1975,nie_thermal_2008}. 
The performance is measured in terms of MLUPS (million lattice updates per second). For the two 13-moment codes, the benchmark is the NACA0012 airfoil of Section~\ref{sec:numerical-tests} with a $2048 \times 2048$ resolution (i.e., $C=256$). The polynomial code was implemented in the Palabos library~\cite{LATT_UnderReview_2019}, and executed for a generic, subsonic flow problem with the same number of nodes. The CPU hardware is an Intel Core i7-8700 CPU at 3.2 GHz with 6 cores, where all 6 cores are exploited through OpenMP (13-moment code) or MPI (polynomial code). The GPU code runs on a single Nvidia GeForce GTX 1080 Ti with 11Gb of central memory. The results are presented in the following table:

\begin{center}
    \begin{tabular}{l|l|l}
         13-moment (CPU) & 13-moment (GPU) & Polynomial (CPU)\\\hline
         0.67 MLUPS & 16.2 MLUPS & 3.55 MLUPS\\
    \end{tabular}
\end{center}

The table shows that the 13-moment CPU code exhibits only a 5-fold speed-down with respect to the fourth-order polynomial code, a price that is cheap to pay for the gain of proper supersonic physics. In addition, the 13-moment approach leads to CPU performances that are typical of standard NSF solvers~\cite{MANOHA_AIAA_2846_2015}, which further confirms the viability of the proposed approach for everyday use.
Interestingly, a more than 20-fold speedup, compared to a multi-core CPU implementation, is obtained with help of a relatively straightforward GPU code, providing more than satisfying performance for the simulation of realistic configurations in an industrial context.
To give an example, the NACA0012 benchmark of Section~\ref{sec:numerical-tests} (i.e., $C=350$) was fully carried out on a single GPU and reached the full duration of 15,000 iterations in less than four hours.


\section{Discussion and outlook \label{sec:conclusion}}
A new type of quadrature free LBM was introduced in the context of compressible flow simulation. By relying on an extended equilibrium introduced, that reproduces the first 13 moments of the Maxwell-Boltzmann distribution, the correct macroscopic behavior was recovered using a lattice composed of as few as 39 discrete velocities in 3D. This is a tremendous improvement as compared to previous similar methods that relied on 343 velocities. Also, the computational cost of the proposed solution procedure is shown to remain in acceptable limits, with a speed down of a factor 5 only compared to a method with polynomial equilibrium (which cannot reproduce the same compressible physics). Nevertheless, the computational overhead can be compensated as the method is shown to allow substantial speed ups through the use of accelerators, as shown through a 20-fold speedup using an Nvidia GPU. This guarantees the viability of the 13-moment approach in an industrial context.

In addition, we proposed a kinetic sensor which is used to locally increase the viscosity accordingly to the departure from equilibrium. Considering a extremely simple formulation of this sensor, stable and accurate simulations were obtained for a 1D Riemann problem, as well as, a high-Reynolds number flow past a NACA0012 airfoil in the supersonic regime.

In view of the drastic improvements obtained with the 13-moment based LBM (as compared to the 5-moment version), we strongly hypothesize that, by simply increasing the number of constraints used for the derivation of the exponential equilibrium, one could extend the present stability range to even higher Mach numbers, including potentially hypersonic conditions. In addition, by considering constraints based on other types of equilibria, this LBM could be extended to simulate other types of flows (multiphase, magnetohydrodynamic, semi-classical, relativistic, etc) and physics (electromagnetism, quantum systems, etc), as it was done for example by DVMs for the simulation of radiative phenomena~\cite{dubroca1999theoretical,charrier2004discrete}. In the context of LBMs, this would be particularly suitable for semi-classical~\cite{coelho_lattice_2018} and relativistic~\cite{gabbana_towards_2017} LBMs, where a large number of constraints must be satisfied to recover the macroscopic behavior of interest, and this imposes the use of very large lattices.

Regarding the kinetic sensor, it only depends on the departure from equilibrium, and can correspondingly be extended to any type of physics. Furthermore, it can also be used to identify under-resolved areas where either a stabilization technique (subgrid scale models and/or shock capturing approach) or mesh refinement are required. For the former, the sensor would definitely offer a better control on the different scales impacted by them -- which is usually the main flaw of these dynamic stabilization techniques~\cite{PIROZZOLI_ARFM_43_2011}.

Eventually, since the current methodology is not tied to any velocity discretization, it would be interesting to apply it to other lattices (D3Q19, D3Q27, D3Q33, D3Q41, etc). This would likely lead to different stability ranges as well as levels of accuracy. In addition, the root-finding algorithm also have a non-negligible impact on the robustness and accuracy of LBMs based on the exponential equilibrium. Hence, it seems important to study in more details the influence of (1) the type of algorithm (gradient based versus gradient free) as well as (2) its convergence criterion. Both investigations will be presented in a forthcoming paper.


\ack{Fruitful discussions with Dr. Orestis Malaspinas and Dr. Dimitrios Kontaxakis are thankfully acknowledged.}

\aucontribute{JL supervised the research and contributed to it, designed the CPU solver, and wrote parts of the manuscript. CC contributed to the research, embedded the proposed model in a wider theoretical context, set up the test cases, and wrote parts of the manuscript. AP authored initial versions of the proposed model, conducted preliminary validation tests, and revised the manuscript. JB wrote and executed the GPU solver, and revised the manuscript. All authors read and approved the manuscript. }

\competing{The authors declare that they have no competing interests.}

\dataccess{This manuscript has no further supporting data.}

\bibliography{bib13}

\begin{thebibliography}{10}
\expandafter\ifx\csname urlstyle\endcsname\relax
  \providecommand{\doi}[1]{(doi:\discretionary{}{}{}#1)}\else
  \providecommand{\doi}{(doi:\discretionary{}{}{}\begingroup
  \urlstyle{rm}\Url)}\fi

\bibitem{he1997priori}
He X, Luo LS. 1997 A priori derivation of the lattice boltzmann equation.
\newblock \emph{Physical Review E} \textbf{55}, R6333.

\bibitem{shan1998discretization}
Shan X, He X. 1998 Discretization of the velocity space in the solution of the
  boltzmann equation.
\newblock \emph{Physical Review Letters} \textbf{80}, 65.

\bibitem{guo2013lattice}
Guo Z, Shu C. 2013 \emph{Lattice Boltzmann method and its applications in
  engineering}, volume~3.
\newblock World Scientific.

\bibitem{kruger_lattice_2017}
Kr{\"u}ger T, Kusumaatmaja H, Kuzmin A, Shardt O, Silva G, Viggen EM. 2017
  \emph{The {Lattice} {Boltzmann} {Method}: {Principles} and {Practice}}.
\newblock Graduate {Texts} in {Physics}. Cham: Springer International
  Publishing.

\bibitem{succi_lattice_2018}
Succi S. 2018 \emph{The {Lattice} {Boltzmann} {Equation} {For} {Complex}
  {States} of {Flowing} {Matter}}.
\newblock Oxford University Press.

\bibitem{leclaire_generalized_2017}
Leclaire S, Parmigiani A, Malaspinas O, Chopard B, Latt J. 2017 Generalized
  three-dimensional lattice {Boltzmann} color-gradient method for immiscible
  two-phase pore-scale imbibition and drainage in porous media.
\newblock \emph{Phys. Rev. E} \textbf{95}, 033306.

\bibitem{parmigiani_lattice_2013}
Parmigiani A, Latt J, Begacem MB, Chopard B. 2013 A lattice {Boltzmann}
  simulation of the {Rhone} river.
\newblock \emph{Int. J. Mod. Phys. C} \textbf{24}, 1340008.

\bibitem{thorimbert_lattice_2018}
Thorimbert Y, Marson F, Parmigiani A, Chopard B, L{\"a}tt J. 2018 Lattice
  {Boltzmann} simulation of dense rigid spherical particle suspensions using
  immersed boundary method.
\newblock \emph{Computers \& Fluids} \textbf{166}, 286--294.

\bibitem{li_application_2018}
Li S, Latt J, Chopard B. 2018 The application of the screen-model based
  approach for stents in cerebral aneurysms.
\newblock \emph{Computers \& Fluids} \textbf{172}, 651--660.

\bibitem{chen_lattice_1998}
Chen S, Doolen GD. 1998 Lattice {Boltzmann} {Method} for {Fluid} {Flows}.
\newblock \emph{Annu. Rev. Fluid Mech.} \textbf{30}, 329--364.

\bibitem{huang_statistical_1987}
Huang K. 1987 \emph{Statistical {Mechanics}, 2nd {Edition}}.
\newblock Wiley, 2 edition.

\bibitem{shan_kinetic_2006}
Shan X, Yuan XF, Chen H. 2006 Kinetic theory representation of hydrodynamics: a
  way beyond the {Navier}{\textendash}{Stokes} equation.
\newblock \emph{J. Fluid Mech.} \textbf{550}, 413.

\bibitem{guo_lattice_2013}
Guo Z, Shu C. 2013 \emph{Lattice {Boltzmann} method and its applications in
  engineering}.
\newblock World Scientific.

\bibitem{siebert_lattice_2008}
Siebert DN, Hegele LA, Philippi PC. 2008 Lattice {Boltzmann} equation linear
  stability analysis: {Thermal} and athermal models.
\newblock \emph{Phys. Rev. E} \textbf{77}, 026707.

\bibitem{coreixas_recursive_2017}
Coreixas C, Wissocq G, Puigt G, Boussuge JF, Sagaut P. 2017 Recursive
  regularization step for high-order lattice {Boltzmann} methods.
\newblock \emph{Phys. Rev. E} \textbf{96}, 033306.

\bibitem{li_temperature-scaled_2019}
Li X, Shi Y, Shan X. 2019 Temperature-scaled collision process for the
  high-order lattice {Boltzmann} model.
\newblock \emph{Phys. Rev. E} \textbf{100}, 013301.

\bibitem{li_coupled_2007}
Li Q, He YL, Wang Y, Tao WQ. 2007 Coupled double-distribution-function lattice
  {Boltzmann} method for the compressible {Navier}-{Stokes} equations.
\newblock \emph{Phys. Rev. E} \textbf{76}, 056705.

\bibitem{feng_compressible_2016}
Feng Y, Sagaut P, Tao WQ. 2016 A compressible lattice {Boltzmann} finite volume
  model for high subsonic and transonic flows on regular lattices.
\newblock \emph{Computers \& Fluids} \textbf{131}, 45--55.

\bibitem{nie_lattice-boltzmann_2009}
Nie X, Shan X, Chen H. 2009 A {Lattice}-{Boltzmann} / {Finite}-{Difference}
  {Hybrid} {Simulation} of {Transonic} {Flow}.
\newblock In \emph{47th {AIAA} {Aerospace} {Sciences} {Meeting} including {The}
  {New} {Horizons} {Forum} and {Aerospace} {Exposition}}, Aerospace {Sciences}
  {Meetings}. American Institute of Aeronautics and Astronautics.

\bibitem{fares_validation_2014}
Fares E, Wessels M, Zhang R, Sun C, Gopalaswamy N, Roberts P, Hoch J, Chen H.
  2014 Validation of a {Lattice}-{Boltzmann} {Approach} for {Transonic} and
  {Supersonic} {Flow} {Simulations}.
\newblock In \emph{52nd {Aerospace} {Sciences} {Meeting}}. National Harbor,
  Maryland: American Institute of Aeronautics and Astronautics.

\bibitem{renard2019improved}
Renard F, Boussuge JF, Feng Y, Sagaut P Improved compressible hybrid lattice
  boltzmann method on standard lattice for subsonic and supersonic flows .Under
  review

\bibitem{gatignol1975theorie}
Gatignol R. 1975 Theorie cinetique de gaz a repartition discrete de vitesses.
\newblock \emph{Lecture Notes in Phys.} \textbf{36}.

\bibitem{charrier1998discrete}
Charrier P, Dubroca B, Feugeas J, Mieussens L. 1998 Discrete-velocity models
  for kinetic nonequilibrium flows.
\newblock \emph{COMPTES RENDUS DE L ACADEMIE DES SCIENCES SERIE I-MATHEMATIQUE}
  \textbf{326}, 1347--1352.

\bibitem{mieussens_discrete_2000}
Mieussens L. 2000 Discrete velocity model and implicit scheme for the bgk
  equation of rarefied gas dynamics.
\newblock \emph{Math. Models Methods Appl. Sci.} \textbf{10}, 1121--1149.

\bibitem{mieussens2001convergence}
Mieussens L. 2001 Convergence of a discrete-velocity model for the
  boltzmann-bgk equation.
\newblock \emph{Computers \& Mathematics with Applications} \textbf{41},
  83--96.

\bibitem{dubroca2001conservative}
Dubroca B, Mieussens L. 2001 A conservative and entropic discrete-velocity
  model for rarefied polyatomic gases.
\newblock In \emph{ESAIM: Proceedings}, volume~10, pp. 127--139. EDP Sciences.

\bibitem{mieussens2004numerical}
Mieussens L, Struchtrup H. 2004 Numerical comparison of bhatnagar--gross--krook
  models with proper prandtl number.
\newblock \emph{Physics of Fluids} \textbf{16}, 2797--2813.

\bibitem{titarev2012efficient}
Titarev V. 2012 Efficient deterministic modelling of three-dimensional rarefied
  gas flows.
\newblock \emph{Communications in Computational Physics} \textbf{12}, 162--192.

\bibitem{baranger2014locally}
Baranger C, Claudel J, H{\'e}rouard N, Mieussens L. 2014 Locally refined
  discrete velocity grids for stationary rarefied flow simulations.
\newblock \emph{Journal of Computational Physics} \textbf{257}, 572--593.

\bibitem{dubroca1999theoretical}
Dubroca B, Feugeas JL. 1999 Theoretical and numerical study on a moment closure
  hierarchy for the radiative transfer equation.
\newblock \emph{Comptes Rendus de l'Academie des Sciences Series I Mathematics}
  \textbf{329}, 915--920.

\bibitem{charrier2004discrete}
Charrier P, Dubroca B, Mieussens L, Turpault R. 2004 Discrete-velocity models
  for numerical simulations in transitional regime for rarefied flows and
  radiative transfer.
\newblock In \emph{Transport in Transition Regimes}, pp. 85--101. Springer.

\bibitem{kogan_derivation_1965}
Kogan AM. 1965 Derivation of {Grad}'s type equations and study of their
  relaxation properties by the method of maximization of entropy.
\newblock \emph{Journal of Applied Mathematics and Mechanics} \textbf{29},
  130--142.

\bibitem{levermore1996moment}
Levermore CD. 1996 Moment closure hierarchies for kinetic theories.
\newblock \emph{Journal of statistical Physics} \textbf{83}, 1021--1065.

\bibitem{presse_principles_2013}
Press{\'e} S, Ghosh K, Lee J, Dill KA. 2013 Principles of maximum entropy and
  maximum caliber in statistical physics.
\newblock \emph{Rev. Mod. Phys.} \textbf{85}, 1115--1141.

\bibitem{frapolli_entropic_2015}
Frapolli N, Chikatamarla SS, Karlin IV. 2015 Entropic lattice {Boltzmann} model
  for compressible flows.
\newblock \emph{Phys. Rev. E} \textbf{92}, 061301.

\bibitem{frapolli_entropic_2016}
Frapolli N, Chikatamarla SS, Karlin IV. 2016 Entropic lattice {Boltzmann} model
  for gas dynamics: {Theory}, boundary conditions, and implementation.
\newblock \emph{Phys. Rev. E} \textbf{93}, 063302.

\bibitem{frapolli_entropic_2017}
Frapolli N. 2017 \emph{Entropic lattice {Boltzmann} models for thermal and
  compressible flows}.
\newblock Ph.D. thesis, ETH Zurich.

\bibitem{BHATNAGAR_PR_94_1954}
Bhatnagar P, Gross E, Krook M. 1954 A model for collision processes in gases.
  {I}. {S}mall amplitude processes in charged and neutral one-component
  systems.
\newblock \emph{Phys. Rev.} \textbf{94}, 511--525.

\bibitem{struchtrup_regularization_2003}
Struchtrup H, Torrilhon M. 2003 Regularization of {Grad}{\textquoteright}s 13
  moment equations: {Derivation} and linear analysis.
\newblock \emph{Physics of Fluids} \textbf{15}, 2668--2680.

\bibitem{rykov_model_1975}
Rykov VA. 1975 A model kinetic equation for a gas with rotational degrees of
  freedom.
\newblock \emph{Fluid Dyn} \textbf{10}, 959--966.

\bibitem{nie_thermal_2008}
Nie X, Shan X, Chen H. 2008 Thermal lattice {Boltzmann} model for gases with
  internal degrees of freedom.
\newblock \emph{Phys. Rev. E} \textbf{77}, 035701.

\bibitem{press_numerical_2007}
Press WH, Teukolsky SA, Vetterling WT, Flannery BP. 2007 \emph{Numerical
  {Recipes}: {The} {Art} of {Scientific} {Computing}}.
\newblock Cambridge University Press, 3rd edition edition.

\bibitem{chapman_mathematical_1990}
Chapman S, Cowling TG, Burnett D. 1990 \emph{The mathematical theory of
  non-uniform gases: an account of the kinetic theory of viscosity, thermal
  conduction and diffusion in gases}.
\newblock Cambridge university press.

\bibitem{thorimbert_automatic_2019}
Thorimbert Y, Lagrava D, Malaspinas O, Chopard B, Latt J. 2019 Automatic {Grid}
  {Refinement} {Criterion} for {Lattice} {Boltzmann} {Method}.
\newblock Submitted for publication, Universit{\'e} de Gen{\`e}ve.

\bibitem{karlin_gibbs_2014}
Karlin IV, B{\"o}sch F, Chikatamarla SS. 2014 Gibbs' principle for the
  lattice-kinetic theory of fluid dynamics.
\newblock \emph{Phys. Rev. E} \textbf{90}, 031302.

\bibitem{BROWNLEE_PRE_74_2006}
Brownlee RA, Gorban AN, Levesley J. 2006 Stabilization of the lattice
  {B}oltzmann method using the ehrenfests' coarse-graining idea.
\newblock \emph{Phys. Rev. E} \textbf{74}, 037703.

\bibitem{BROWNLEE_PRE_75_2007}
Brownlee RA, Gorban AN, Levesley J. 2007 Stability and stabilization of the
  lattice {B}oltzmann method.
\newblock \emph{Phys. Rev. E} \textbf{75}, 036711.

\bibitem{BROWNLEE_PA_387_2008}
Brownlee R, Gorban A, Levesley J. 2008 Nonequilibrium entropy limiters in
  lattice {B}oltzmann methods.
\newblock \emph{Physica A} \textbf{387}, 385 -- 406.

\bibitem{gorban_enhancement_2014}
Gorban AN, Packwood DJ. 2014 Enhancement of the stability of lattice
  {Boltzmann} methods by dissipation control.
\newblock \emph{Physica A: Statistical Mechanics and its Applications}
  \textbf{414}, 285--299.

\bibitem{sod_survey_1978}
Sod GA. 1978 A survey of several finite difference methods for systems of
  nonlinear hyperbolic conservation laws.
\newblock \emph{Journal of Computational Physics} \textbf{27}, 1--31.

\bibitem{coreixas_high-order_2018}
Coreixas CG. 2018 \emph{High-order extension of the recursive regularized
  lattice {Boltzmann} method}.
\newblock phd, CERFACS.

\bibitem{hafez_simulations_2007}
Hafez M, Wahba E. 2007 Simulations of viscous transonic flows over lifting
  airfoils and wings.
\newblock \emph{Computers \& Fluids} \textbf{36}, 39--52.

\bibitem{KRUGER_Book_2017}
Kr\"{u}ger T, Kusumaatmaja H, Kuzmin A, Shardt O, Silva G, Viggen EM. 2017
  \emph{The Lattice {B}oltzmann Method: Principles and Practice}.
\newblock Springer International Publishing.

\bibitem{coreixas2019comprehensive}
Coreixas C, Chopard B, Latt J. 2019 Comprehensive comparison of collision
  models in the lattice boltzmann framework: Theoretical investigations.
\newblock \emph{Phys. Rev. E} \textbf{100}, 033305.

\bibitem{SHAN_PRE_100_2019}
Shan X. 2019 Central-moment-based galilean-invariant multiple-relaxation-time
  collision model.
\newblock \emph{Phys. Rev. E} \textbf{100}, 043308.

\bibitem{SHAN_JCS_17_2016}
Shan X. 2016 The mathematical structure of the lattices of the lattice
  {B}oltzmann method.
\newblock \emph{J. Comput. Sci.} \textbf{17}, 475 -- 481.

\bibitem{LATT_UnderReview_2019}
Latt J, Malaspinas O, Kontaxakis D, Parmigiani A, Lagrava D, Brogi F, Belgacem
  MB, Thorimbert Y, Leclaire S, Li S, Kotsalos C, Marson F, Coreixas C,
  Petkantchin R, Raynaud F, Conradin R, Beny J, Chopard B. 2019 Palabos:
  Parallel lattice boltzmann solver .(under review)

\bibitem{MANOHA_AIAA_2846_2015}
Manoha E, Caruelle B. 2015 Summary of the lagoon solutions from the benchmark
  problems for airframe noise computations-iii workshop.
\newblock In \emph{21st AIAA/CEAS Aeroacoustics Conference}, p. 2846.

\bibitem{coelho_lattice_2018}
Coelho RCV, Doria MM. 2018 Lattice {Boltzmann} method for semiclassical fluids.
\newblock \emph{Computers \& Fluids} \textbf{165}, 144--159.

\bibitem{gabbana_towards_2017}
Gabbana A, Mendoza M, Succi S, Tripiccione R. 2017 Towards a unified lattice
  kinetic scheme for relativistic hydrodynamics.
\newblock \emph{Phys. Rev. E} \textbf{95}, 053304.

\bibitem{PIROZZOLI_ARFM_43_2011}
Pirozzoli S. 2011 Numerical methods for high-speed flows.
\newblock \emph{Annu. Rev. Fluid Mech} \textbf{43}, 163--194.

\end{thebibliography}
\bibliographystyle{rsta}

\end{document}